\documentclass[aps,prx,twocolumn,10pt,showpacs,superscriptaddress,amsmath,amssymb]{revtex4-1}

\usepackage{graphicx}

\begin{document}

\title{Unified view of quantum amplification based on quantum transformation}
\author{Meng-Jun Hu}
\author{Yong-Sheng Zhang}\email{yshzhang@ustc.edu.cn}
\affiliation{Key Laboratory of Quantum Information, University of Science and Technology of China, Hefei 230026, China}
\affiliation{ Synergetic Innovation Center of Quantum Information and Quantum Physics, University of Science and Technology of China, Hefei 230026, China}

\date{\today}

\begin{abstract}
A general framework of quantum state amplification using the language of quantum state transformation is given systematically for the first time. The concept of amplification of quantum states is defined specifically and the amplification of a set of quantum states is formulated generally as the transformation of quantum states. Three different kinds of important quantum amplifications, i.e., deterministic noisy quantum amplification, probabilistic noiseless quantum amplification, and deterministic noiseless quantum amplification are identified and discussed. For deterministic quantum amplification, the linearity of amplification is proven to be incompatible with the noiseless amplification while it is not true for probabilistic quantum amplification. However, deterministic noiseless quantum amplification is shown physically attainable if the linearity of amplification is given up. The connection between the gain of amplification and the successful probability is discussed for probabilistic quantum amplification. Assuming that successful probability is the same for all quantum states to be amplified, we obtain a generally valid relation between the gain of amplification and the successful probability. Particular interest is given to phase-preserving quantum amplification of Gaussian states which has been shown of theoretical interest and of practical importance in quantum information and quantum communication recently. Our results of quantum state amplification not only enrich the research of quantum amplification but also can be helpful for further practical applications.  
\end{abstract}

\pacs{03.67.-a, 42.50.Dv, 42.50.Ex}

\maketitle

\section{\bf Introduction}
Quantum amplification (QA) is at the heart of quantum measurement and quantum metrology. Restricted by fundamental laws of physics, it is difficult for detectors to measure sufficiently weak signal, especially for quantum signal. Amplifying signal first and then detect it with proper detector thus becomes a general way for signal detection. Unfortunately, the intrinsic noise accompanying the signal is also amplified during the process of amplification and extra noises may be introduced to make the signal-to-noise ratio (SNR) worse. 

For a linear phase-preserving quantum amplifier, due to the constraint of bosonic commutation relation $[a,a^{\dagger}]=1$, it has been shown that there is at least $(g^{2}-1)\hbar\omega$ total noise power per unit bandwidth out of its output-port, where $g^{2}$ is the power gain \cite{1,2}. As quantum signal amplification is equivalent to quantum state amplification, it seems that a universal linear phase-preserving quantum amplifier that can amplify any coherent state determinately and noiselessly is impossible \cite{3}. For the special set of coherent states, however, the probabilistic noiseless amplification of coherent states is physically realizable \cite{4,5,6} and there have been some experimental reports on the realization of noiseless amplification of quantum light states \cite{7,8,9}. Moreover, regarding quantum state amplification as quantum state transformation, our recent research shows that for a certain set of coherent states, which satisfies particular conditions, there always exists a specific quantum amplifier that can amplify coherent state in that set determinately and noiselessly \cite{10}. 

Although various investigations are given to quantum amplification, there is still lack of a general framework to discuss it. In this paper, a general framework of quantum state amplification using the language of quantum state transformation is given systematically for the first time. The paper is organized as follows. In Sec. II, the concept of amplification of quantum state is definitely defined. The general discussion of amplification of quantum states using the language of quantum state transformation is given and three different important amplifications, i.e., deterministic noisy QA, probabilistic noiseless QA and deterministic noiseless QA are identified and discussed. In Sec. III, phase-preserving QA is considered and the particular interest is given to the deterministic noiseless Gaussian state amplification. An example of application in phase measurement using deterministic noiseless QA is shown and the situation of amplified states in noisy environment is discussed. In Sec. IV, we give a summary of these results.
\section{\bf Amplification of quantum states}
Before the detail discussion of quantum state amplification, an explicit definition of amplification of quantum state is necessary. The quantum state amplification defined as quantum state transformation is then illustrated and three different kinds of important QA are identified and discussed.
\subsection{\bf The definition of amplification of quantum state}
Suppose a system in which we have interest is in quantum state $|\psi\rangle$. When we measure the observable $\hat{A}$ in this system, the physical quantity obtained in the measurement is the expectation of $\hat{A}$ in state $|\psi\rangle$, that is $\langle\psi|\hat{A}|\psi\rangle $. The expectation value $\langle\psi|\hat{A}|\psi\rangle$ of observable $\hat{A}$ in state $|\psi\rangle$ is usually called quantum signal. Amplification of the quantum signal means the amplification of the expectation value $\langle\psi|\hat{A}|\psi\rangle$, which is equivalent to the amplification of state $|\psi\rangle$. Now suppose that the system evolves under an operation that transforms the state $|\psi\rangle$ into the another state $|\phi\rangle$. If the expectation value of observable $\hat{A}$ in the state $|\phi\rangle$ is larger than the expectation value in state $|\psi\rangle$, that is $\langle\phi|\hat{A}|\phi\rangle>\langle\psi|\hat{A}|\psi\rangle$, then we say the quantum signal is amplified, or equivalently, the state is amplified with respect to observable $\hat{A}$. The process of quantum signal amplification is thus actually a process of quantum state transformation. Note that QA is related with the measurement of observable. We thus give the following definition of amplification of quantum state.

{\textbf{Definition 1:}}
If there exists an operation that transforms the state $|\psi\rangle$ of quantum system into the another state $|\phi\rangle$ such that $\langle\phi|\hat{A}|\phi\rangle>\langle\psi|\hat{A}|\psi\rangle$, then this process is called the amplification of the state $|\psi\rangle$ with respect to observable $\hat{A}$. The state $|\phi\rangle$ is called the amplified state of the state $|\psi\rangle$ with respect to observable $\hat{A}$.

If $\langle\phi|\hat{A}|\phi\rangle=g\langle\psi|\hat{A}|\psi\rangle$ with $g>1$, we say that the state $|\psi\rangle$ is amplified with gain of $g$. If there exists a setup that can amplify state $|\psi\rangle$ into the amplified state $|\phi\rangle$ with the gain of $g$ , then we call that setup a quantum amplifier with gain of $g$ of the amplification. A quantum amplifier is thus a physical system whose function is to transform the input state $|\psi\rangle$ to the amplified state $|\phi\rangle$. 

Before the amplification, the fluctuation of observable $\hat{A}$ in state $|\psi\rangle$ is $\Delta\hat{A}_{|\psi\rangle}=\sqrt{\langle\psi|\hat{A}^{2}|\psi\rangle-\langle\psi|\hat{A}|\psi\rangle^{2}}$. After the amplification, the fluctuation of observable $\hat{A}$ in the amplified state $|\phi\rangle$ becomes $\Delta\hat{A}_{|\phi\rangle}=\sqrt{\langle\phi|\hat{A}^{2}|\phi\rangle-\langle\phi|\hat{A}|\phi\rangle^{2}}$. The amplification is called noiseless if $\Delta\hat{A}_{|\phi\rangle}=\Delta\hat{A}_{|\psi\rangle}$, otherwise, the amplification is noisy. The definition of noiseless amplification of quantum state is thus defined as below.


{\textbf{Definition 2:}}
The quantum amplification is called noiseless if the fluctuation of observable $\hat{A}$ is the same in the state $|\psi\rangle$ and the amplified state $|\phi\rangle$, otherwise the quantum amplification is  noisy.

The above definition can be easily extended to mixed state of a quantum system. Suppose that the state of the quantum system is $\rho$, if we send this state to a quantum amplifier with gain of $g$ and the amplified state is $\sigma$, then we have $\langle\hat{A}\rangle_{\sigma}=g\langle\hat{A}\rangle_{\rho}$, where $\langle\hat{A}\rangle_{\sigma}=\mathrm{Tr}(\hat{A}\sigma)$ and $\langle\hat{A}\rangle_{\rho}=\mathrm{Tr}(\hat{A}\rho)$ are the expectation values of observable $\hat{A}$ in state $\sigma$ and $\rho$ respectively. The amplification is noiseless if $\Delta\hat{A}_{\sigma}=\Delta\hat{A}_{\rho}$, where $\Delta\hat{A}_{\sigma}=\sqrt{\mathrm{Tr}(\hat{A}^{2}\sigma)-\mathrm{Tr}(\hat{A}\sigma)^{2}}$ and $\Delta\hat{A}_{\rho}=\sqrt{\mathrm{Tr}(\hat{A}^{2}\rho)-\mathrm{Tr}(\hat{A}\rho)^{2}}$ are fluctuations of observable $\hat{A}$ in  state $\sigma$ and $\rho$ respectively.

The definition of amplification of quantum state with respect to two observables or more than two observables is also valid. For example, in the case of two observables $\hat{A}$ and $\hat{B}$, we have $\langle\phi|\hat{A}|\psi\rangle=g_{1}\langle\psi|\hat{A}|\psi\rangle$ and $\langle\phi|\hat{B}|\phi\rangle=g_{2}\langle\psi|\hat{B}|\psi\rangle$, where $g_{1}$ and $g_{2}$ are the gains of amplification with respect to observables $\hat{A}$ and $\hat{B}$ respectively. The gains of $g_{1}$ and $g_{2}$ can be equal to each other that $g_{1}=g_{2}$ or unequal to each other that $g_{1}\neq g_{2}$. The amplification is noiseless if and only if $\Delta\hat{A}_{|\phi\rangle}=\Delta\hat{A}_{|\psi\rangle}$ and $\Delta\hat{B}_{|\phi\rangle}=\Delta\hat{B}_{|\psi\rangle}$. If $\Delta\hat{A}_{|\phi\rangle}=\Delta\hat{A}_{|\psi\rangle}$ but $\Delta\hat{B}_{|\phi\rangle}\neq\Delta\hat{B}_{|\psi\rangle}$, then the amplification is only noiseless with respect to observable $\hat{A}$. Conversely, if $\Delta\hat{A}_{|\phi\rangle}\neq\Delta\hat{A}_{|\psi\rangle}$ but $\Delta\hat{B}_{|\phi\rangle}=\Delta\hat{B}_{|\psi\rangle}$, then the amplification is only noiseless with respect to observable $\hat{B}$. In all other cases, the amplification is not noiseless.

\subsection{\bf Quantum state amplification viewed as quantum state transformation}
According to the definition of amplification of quantum state, the quantum state amplification is actually a quantum state transformation, thus it is natural to discuss the amplification of the quantum state using the language of quantum state transformation. In fact, the language of quantum state transformation provides a unified framework describing the quantum state cloning, the unambiguous discrimination of states, and the quantum state amplification \citep{11,12,13}.

In the quantum operation theory, any physically permissible state transformation of a quantum system can be determined by a completely positive (CP), linear, trace non-increasing map: $\Lambda: \rho\rightarrow\Lambda(\rho)$ \citep{14,15}. The transformation is physically realizable in principle if such a map $\Lambda$ mathematically exists.

The first representation theorem gives the general representation of the CP, linear, trace nonincreasing map $\Lambda$ \citep{16}. It states that any CP, linear, trace non-increasing map $\Lambda$ can be represented by an operator-sum form $\Lambda(\rho)=\sum_{k}\hat{M}_{k}\rho\hat{M}_{k}^{\dagger}$, where $\hat{M}_{k}$ is the Kraus operator that satisfies $\sum_{k}\hat{M}_{k}^{\dagger}\hat{M}_{k}\leq\hat{I}$. For deterministic transformation $\sum_{k}\hat{M}_{k}^{\dagger}\hat{M}_{k}=\hat{I}$, while for probabilistic transformation $\sum_{k}\hat{M}_{k}^{\dagger}\hat{M}_{k}<\hat{I}$ \citep{17,18}. The unitary evolution of a quantum system that $\Lambda(\rho)=\hat{U}\rho\hat{U}^{\dagger}$ is a special case of this representation.
Another more physical way to represent the map $\Lambda$ is considering the unitary evolution on an enlarged Hilbert space. If an ancillary system is introduced to a quantum system, and a unitary transformation is applied onto the composite system, the map $\Lambda$ is realized when we make projective measurement on the ancillary system. Mathematically, it can be expressed as $\Lambda(\rho)=\mathrm{Tr}_{E^{\prime}}\lbrace \hat{U}\rho\bigotimes\rho_{E}\hat{U}^{\dagger}\hat{I}\bigotimes P_{E^{\prime}}\rbrace $, where $\rho_{E}$ is the initial state of the ancillary system and $P_{E^{\prime}}$ is a projector in the transformed ancillary Hilbert space \citep{15}. For convenience, we will use both of them optionally. 

In practice, the significant case is the amplification of a set of quantum states. The amplification of a specific state is trivial since we can always prepare the system in the expected quantum state.  
For simplicity, the amplification of a set of $N$ linear-independent pure states $\Pi\equiv\lbrace|\psi_{i}\rangle\rbrace$ with $i=1,2...N$ is preferred to be considered.
Suppose that the set of quantum states $\Xi\equiv\lbrace|\phi_{i}\rangle\rbrace$ is the corresponding amplified set of quantum states with respect to observable $\hat{A}$. The operator $\hat{A}$ can be a single observable, or represent a set of specific observables, which will not be mentioned below except for the case of necessary. For any quantum state $|\psi_{i}\rangle$ in the set $\Pi$, there should exist a unitary operation $\hat{U}$  that transforms the state $|\psi_{i}\rangle$ into the corresponding amplified state $|\phi_{i}\rangle$ with the assistance of an ancillary system. In general, the transformation is probabilistic and the successful probability $p$ may be varied for different states in the set $\Pi$. For an arbitrary state $|\psi_{i}\rangle$ in set $\Pi$, we thus have the following formula of quantum state transformation:
\begin{equation}
\hat{U}|\psi_{i}\rangle|0\rangle=\sqrt{p_{i}}|\phi_{i}\rangle|\mu_{i}\rangle|+\rangle+\sqrt{1-p_{i}}|Fail\rangle|\nu_{i}\rangle|-\rangle.
\end{equation}
Here $|+\rangle$, $|-\rangle$ are the pointer states that indicate the amplification is successful or failure respectively. The initial state of the ancillary system is simply denoted by $|0\rangle$ and it is assumed that there is no correlation between the system and the ancillary initially. $|\mu_{i}\rangle$ and $|\nu_{i}\rangle$ are states of the ancillary system corresponding to successful and failure amplification respectively.  

The complex conjugate of Eq. (1) can be written as
\begin{equation}
\langle0|\langle\psi_{j}|\hat{U}^{\dagger}=\sqrt{p_{j}}\langle\phi_{j}|\langle\mu_{j}|\langle+|+\sqrt{1-p_{j}}\langle Fail|\langle\nu_{j}|\langle-|.
\end{equation}
Taking the inner product of Eq. (1) and Eq. (2) gives
\begin{equation}
\langle\psi_{j}|\psi_{i}\rangle=\sqrt{p_{i}p_{j}}\langle\phi_{j}|\phi_{i}\rangle\langle\mu_{j}|\mu_{i}\rangle+\sqrt{(1-p_{i})(1-p_{j})}\langle\nu_{j}|\nu_{i}\rangle.
\end{equation}
Equation (3) is valid for any pair of states $\lbrace|\psi_{i}\rangle,|\psi_{j}\rangle\rbrace$ in the set $\Pi$, and it can be recast as
\begin{equation}
G_{\Pi}=G_{\Xi}\circ \Omega+K.
\end{equation}
Here $G_{\Pi}$ and $G_{\Xi}$ are Gram matrices of set $\Pi$ and $\Xi$ respectively. The Gram matrix $\Omega$ is defined as $\Omega_{ij}=\sqrt{p_{i}p_{j}}\langle\mu_{j}|\mu_{i}\rangle$ and the Gram matrix $K$ is defined as $K_{ij}=\sqrt{(1-p_{i})(1-p_{j})}\langle\nu_{j}|\nu_{i}\rangle$.

It is obvious that the matrix $\Omega$ satisfies the following conditions: $\Omega\geq0$; $\mathrm{diag}(\Omega)=\vec{p}=(p_{1},p_{2},...,p_{N})$; and $G_{\Pi}-G_{\Xi}\circ\Omega\geq0$. According to quantum transformation theorem of sets of pure states \cite{18}, the matrix $\Omega$ can be factorized as $C^{\dagger}C$ where $C=[c_{ki}]$ is a $M\times N$ matrix. The amplification Kraus operators can be constructed as
\begin{equation}
\hat{M}_{ks}=\sum_{i}\dfrac{c_{ki}}{\langle\tilde{\psi_{i}}|\psi_{i}\rangle}|\phi_{i}\rangle\langle\tilde{\psi_{i}}|,
\end{equation}
where $\hat{M}_{ks}(k=1,2,...,M)$ are the Kraus operators for successful amplification. State $|\tilde{\psi_{i}}\rangle$ is orthogonal to any state in set $\Omega$ except for state $|\psi_{i}\rangle$ and $\langle\tilde{\psi_{i}}|\psi_{j}\rangle=\gamma_{i}\delta_{ij}$, where $\gamma_{i}\neq 0$ is a constant.

For the successful amplification, it can be seen that for any state $|\psi_{i}\rangle$
\begin{equation}
\hat{M}_{ks}|\psi_{i}\rangle=c_{ki}|\phi_{i}\rangle.
\end{equation}

Though only pure state amplification is considered above, the extension to the case of general state amplification is not difficult. Suppose that the set of $N$ linear-independent general quantum states $\Pi=\lbrace\rho_{1},\rho_{2},...,\rho_{N}\rbrace$ is to be amplified, and the set $\Xi=\lbrace\sigma_{1},\sigma_{2},...,\sigma_{N}\rbrace$ is the corresponding amplified set of quantum states. For any state $\rho_{i}$ in set $\Pi$, the transformation of amplification can be formulated as:
\begin{equation}
\hat{U}\rho_{i}\otimes\rho_{E}\hat{U}^{\dagger}=p_{i}\sigma_{i}\otimes\tau_{i}\otimes|+\rangle\langle+|+(1-p_{i})\varrho\otimes\upsilon_{i}\otimes|-\rangle\langle-|,
\end{equation}
where $|+\rangle\langle+|,|-\rangle\langle-|$ are pointer states, $\tau_{i}$ and $\upsilon_{i}$ are states of the ancillary system corresponding to successful and failure amplification respectively, and $\varrho$ is the failure state of the quantum system.

Taking the trace of the product of Eq. (7) and its conjugate gives
\begin{equation}
\mathrm{Tr}(\rho_{i}\rho_{j})=p_{i}p_{j}\mathrm{Tr}(\sigma_{i}\sigma_{j})\mathrm{Tr}(\tau_{i}\tau_{j})+(1-p_{i})(1-p_{j})\mathrm{Tr}(\upsilon_{i}\upsilon_{j}),
\end{equation}
which is valid for any pair of states $\lbrace\rho_{i},\rho_{j}\rbrace$ in set $\Pi$. In the above formula, we have assumed that $\rho_{E}$ is a pure state and $\mathrm{Tr}(\varrho^{2})$ is a constant that involved in term $\mathrm{Tr}(\upsilon_{i}\upsilon_{j})$.
The Kraus operator representation of successful amplification now becomes
\begin{equation}
\hat{M}_{ks}\rho_{i}\hat{M}_{ks}^{\dagger}=|c_{ki}|^{2}\sigma_{i}.
\end{equation}

Since a mixed state $\rho$ can be considered as an ensemble of pure quantum states, the amplification of a set of general quantum states  is equivalent to the amplification of a specific set of pure quantum states. In the following discussion, the QA of a set of states refers to the amplification of pure states without pointing out it explicitly.
\subsection{\bf The classification of quantum amplification}
There are mainly two different kinds of quantum amplification, i.e., deterministic QA and probabilistic QA that can be obviously seen from Eq. (8). 

We first consider the deterministic amplification of the set of quantum states $\Pi=\lbrace\rho_{1},\rho_{2},...,\rho_{N}\rbrace$. The successful probability of amplification is unity for any state $\rho_{i}$ in set $\Pi$, that is $p_{i}=1,\forall\rho_{i}\subseteq\Pi$. The second term in the right-hand side (RHS) of Eq. (7) vanishes for this case and Eq. (8) becomes
\begin{equation}
 \mathrm{Tr}(\rho_{i}\rho_{j})=\mathrm{Tr}(\sigma_{i}\sigma_{j})\mathrm{Tr}(\tau_{i}\tau_{j}).
\end{equation} 
Since the overlap of any two states is less than unity, we have $\mathrm{Tr}(\rho_{i}\rho_{j})=\mathrm{Tr}(\sigma_{i}\sigma_{j})\mathrm{Tr}(\tau_{i}\tau_{j})\leq \mathrm{Tr}(\sigma_{i}\sigma_{j})$ for any states $\rho_{i},\rho_{j}\subseteq\Pi$, which implies that the overlap between any two states to be amplified is no more than the overlap between two corresponding amplified states for deterministic amplification. From the point of view of information, the distinguishability of any two states to be amplified does not increase after the deterministic QA. The underlying physics of this consequence is that a quantum system will dissipate some of its information to the environment because of the interaction between them. In fact, for any quantum operation $L$ on the set of states $\Pi$, the distinguishability of any two states $\rho_{i},\rho_{j}$ in the set does not increase after operation, which means
\begin{equation}
D(\rho_{i},\rho_{j})\geq D(L\rho_{i},L\rho_{j}),
\end{equation}
where $D(\rho_{i},\rho_{j})$ represents the distinguishability of states $\rho_{i},\rho_{j}$.
The distinguishability does not increase after amplification provides us an essential tool to investigate deterministic QA further. 

Suppose that the amplification is noiseless, which means that the fluctuation of observable $\hat{A}$ in any state $\rho_{i}\subseteq\Pi$ and its corresponding amplified state $\sigma_{i}\subseteq\Xi$ stay unchanged, i.e., $\Delta\hat{A}_{\rho_{i}}=\Delta\hat{A}_{\sigma_{i}}$. For a linear quantum amplifier with fixed gain $g$ of amplification, which means $\mathrm{Tr}(\hat{A}\sigma_{i})=g\mathrm{Tr}(\hat{A}\rho_{i})$ for any state $\rho_{i}\subseteq\Pi$, we will show that the linearity of quantum amplifier is incompatible with the noiseless amplification.

\begin{figure*}
\centering
\includegraphics[scale=0.35]{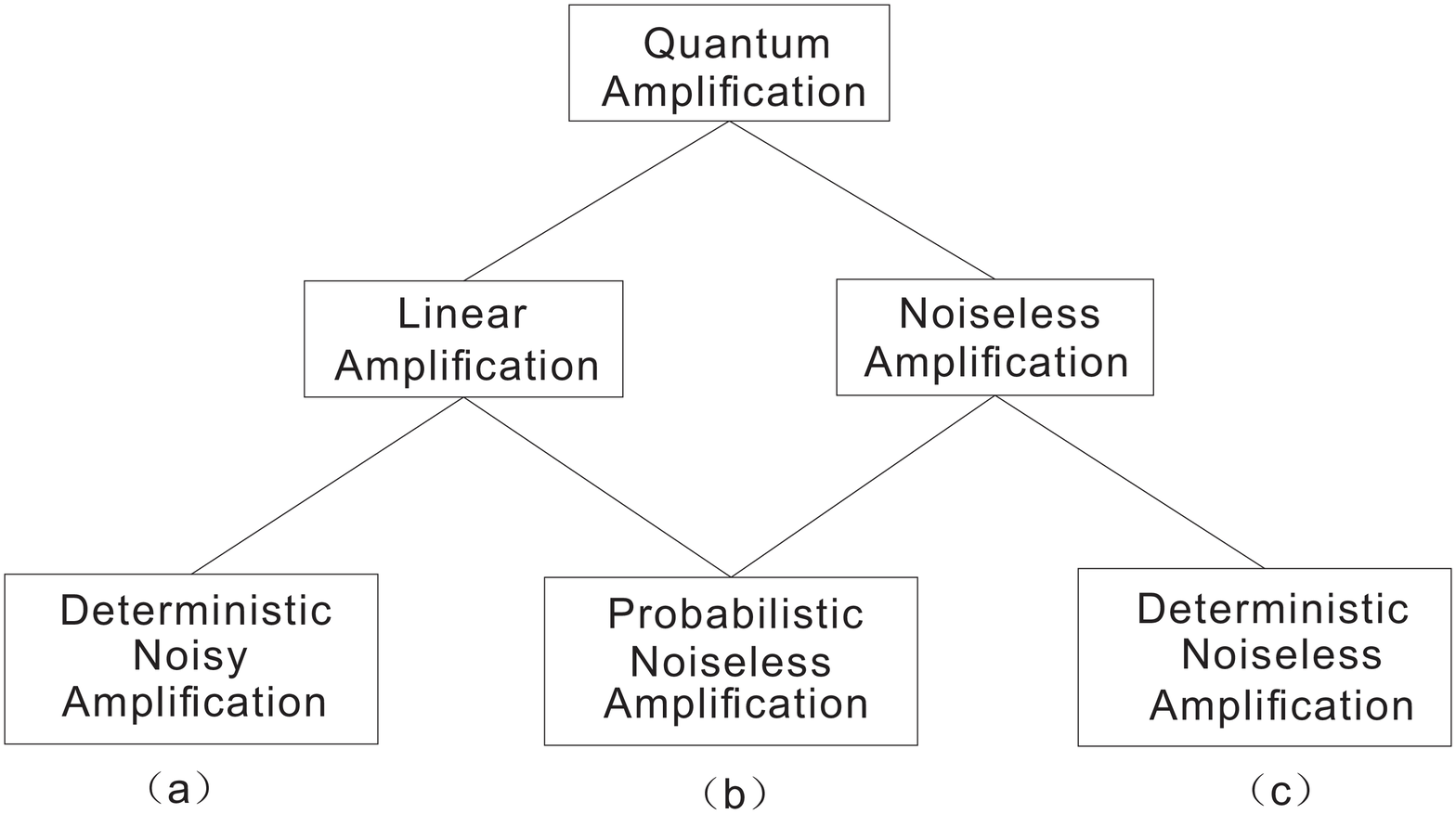}
\caption{{\it The relation of three different kinds of typical quantum amplification.} For deterministic amplification, the linear amplification is incompatible with the noiseless amplification. Deterministic noiseless amplification is only possible for nonlinear quantum amplifier and linear deterministic amplification must be noisy. However, the linear amplification can be noiseless if the amplification is probabilistic. Among the three kinds of quantum amplification, (b) and (c) belong to noiseless amplification and (a) and (b) belong to linear amplification.}
\end{figure*}

Suppose that the eigenvalues of observable $\hat{A}$ are $\lbrace a_{f}\rbrace$ and the corresponding eigenstates are $\lbrace|a_{f}\rangle\rbrace$ so that $\hat{A}=\sum_{f}a_{f}|a_{f}\rangle\langle a_{f}|$. The expectation of observable $\hat{A}$ in any quantum state $\rho$ now can be expressed as
\begin{equation}
\mathrm{Tr}(\hat{A}\rho)=\sum_{f}a_{f}\langle a_{f}|\rho|a_{f}\rangle.
\end{equation}
For a linear quantum amplifier with gain $g$, $\mathrm{Tr}(\hat{A}\rho_{i})=g\mathrm{Tr}(\hat{A}\sigma_{i})$ holds for any state $\rho_{i}\subseteq\Pi$ and the corresponding amplified state $\sigma_{i}\subseteq\Xi$.

In the representation of observable $\hat{A}$, the fluctuation of $\hat{A}$ in quantum state $\rho_{i}$ reads
\begin{equation}
\begin{split}
\Delta\hat{A}_{\rho_{i}}^{2}&=\mathrm{Tr}(\hat{A}^{2}\rho_{i})-\mathrm{Tr}(\hat{A}\rho_{i})^{2} \\
&=\sum_{f}a_{f}^{2}\langle a_{f}|\rho_{i}|a_{f}\rangle-(\sum_{f}a_{f}\langle a_{f}|\rho_{i}|a_{f}\rangle)^{2}.
\end{split}
\end{equation}
In the amplified state $\sigma_{i}$, the fluctuation of observable $\hat{A}$ is
\begin{equation}
\begin{split}
\Delta\hat{A}_{\sigma_{i}}^{2}&=\sum_{f}a_{f}^{2}\langle a_{f}|\sigma_{i}|a_{f}\rangle-(\sum_{f}a_{f}\langle a_{f}|\sigma_{i}|a_{f}\rangle)^{2}\\
&=\sum_{f}a_{f}^{2}\langle a_{f}|\sigma_{i}|a_{f}\rangle-g^{2}(\sum_{f}a_{f}\langle a_{f}|\rho_{i}|a_{f}\rangle)^{2}.
\end{split}
\end{equation}
If the amplification is noiseless, then $\Delta\hat{A}_{\rho_{i}}^{2}=\Delta\hat{A}_{\sigma_{i}}^{2}$ gives
\begin{equation}
(g^{2}-1)\mathrm{Tr}(\hat{A}\rho_{i})^{2}=\mathrm{Tr}\lbrace\hat{A}^{2}(\sigma_{i}-\rho_{i})\rbrace.
\end{equation}
Considering the term in the RHS of Eq.(15), we have
\begin{equation}
\begin{split}
\mathrm{Tr}\lbrace\hat{A}^{2}(\sigma_{i}-\rho_{i})\rbrace=\sum_{f}a_{f}^{2}\langle a_{f}|(\sigma_{i}-\rho_{i})|a_{f}\rangle 
\\
<a_{max}(g-1)\mathrm{Tr}(\hat{A}\rho_{i}),
\end{split}
\end{equation}
where $a_{max}$ is the maximum eigenvalue of observable $\hat{A}$.
Substituting  Eq.(16) into Eq. (15) gives
\begin{equation}
(g^{2}-1)\mathrm{Tr}(\hat{A}\rho_{i})^{2}<a_{max}(g-1)\mathrm{Tr}(\hat{A}\rho_{i}),
\end{equation}
which is obviously not hold in general.

It is obvious from Eq. (14) that $\Delta\hat{A}_{\sigma_{i}}\neq \Delta\hat{A}_{\rho_{i}}$ except for $g=1$, which means there is no amplification at all. We thus conclude that a deterministic linear quantum amplifier with fixed gain $g$ can not be noiseless. 

$\mathrm{Tr}(\rho_{i}\rho_{j})$ is equal to the overlap of two distribution of $\rho_{i}$ and $\rho_{j}$ in the representation of $\hat{A}$, since
\begin{equation}
\mathrm{Tr}(\rho_{i}\rho_{j})=\sum_{f,g}\langle a_{f}|\rho_{i}|a_{g}\rangle\langle a_{g}|\rho_{j}|a_{f}\rangle.
\end{equation}
The sum should be replaced by an integral if $\hat{A}$ is a continuous observable
\begin{equation}
\mathrm{Tr}(\rho_{i}\rho_{j})=\int dxdy\langle x|\rho_{i}|y\rangle\langle y|\rho_{j}|x\rangle.
\end{equation}
The added noise to the amplified states must ensure that the overlap between amplified states larger than corresponding initial states. It is easy to understand just with this added noise the distinguishability of any two amplified quantum states does not increase. 

Although the deterministic amplification can not be noiseless for a linear quantum amplifier, the deterministic noiseless amplification is attainable if the gain of amplification can be state-dependent, i.e., the quantum amplifier is nonlinear \cite{10}. Here, we assume that the gain of amplification of state $\rho_{i}$ is $g_{i}$, i.e., $\mathrm{Tr}(\hat{A}\sigma_{i})=g_{i}\mathrm{Tr}(\hat{A}\rho_{i})$. For noiseless amplification, the overlap between any two states $\rho_{i},\rho_{j}\subseteq\Pi$ can be less than the overlap between two corresponding amplified states, that means $\mathrm{Tr}(\rho_{i}\rho_{j})\leqslant\mathrm{Tr}(\sigma_{i}\sigma_{j})$, which is impossible for a linear quantum amplifier. 
The distinguishability of any two states $\rho_{i},\rho_{j}\subseteq\Pi$ does not increase for a deterministic noiseless amplification means $D(\rho_{i},\rho_{j})\geqslant D(\sigma_{i},\sigma_{j})$  is the only physical requirement. For the set of states $\Pi$ to be amplified and the noiseless amplified set of states $\Xi$, if $D(\rho_{i},\rho_{j})\geqslant D(\sigma_{i},\sigma_{j})$ for any two states $\rho_{i},\rho_{j}$ and their corresponding noiseless amplified states $\sigma_{i},\sigma_{j}$ is satisfied, then such deterministic noiseless quantum amplifier always can be constructed physically.

The linear quantum amplifier with gain $g$ can be noiseless if we do not demand the amplification is deterministic, and in this case the only physical requirement is that Eq. (8) must be hold. The probabilistic noiseless linear quantum amplifier is thus physically allowed. 
The relation between the gain $g$ and the successful probability $p_{i}$ can also be obtained from Eq. (8) since for noiseless amplification $\mathrm{Tr}(\rho_{i}\rho_{j})=f_{ij}(g)\mathrm{Tr}(\sigma_{i}\sigma_{j})$, where $f(g)$ is proportional to $g$ and depends on the states $\rho_{i}$ and $\rho_{j}$ to be amplified. The form of $f_{ij}(g)$ relies on the specific case we consider, for instance, in the noiseless amplification of coherent states, $f_{ij}(g)=e^{(g^{2}-1)|\alpha_{i}-\alpha_{j}|^{2}}$, where $|\alpha_{i}\rangle$ and $|\alpha_{j}\rangle$ are coherent states to be amplified. The Eq. (8) thus becomes
\begin{equation}
f_{ij}(g)=p_{i}p_{j}\mathrm{Tr}(\tau_{i}\tau_{j})+(1-p_{i})(1-p_{j})\dfrac{\mathrm{Tr}(\upsilon_{i}\upsilon_{j})}{\mathrm{Tr}(\sigma_{i}\sigma_{j})}.
\end{equation} 
Assuming that the successful probability is the same for all states $\rho_{i}\subseteq\Pi$ , then we have
\begin{equation}
f_{ij}(g)=C_{i,j}p^{2}+V_{i,j}(1-p)^{2},
\end{equation}
where $C_{i,j}\equiv\mathrm{Tr}(\tau_{i}\tau_{j})$ and $V_{i,j}\equiv \dfrac{\mathrm{Tr}(\upsilon_{i}\upsilon_{j})}{\mathrm{Tr}(\sigma_{i}\sigma_{j})}$. Eq. (21) is valid for any pair of states $\rho_{i},\rho_{j}\subseteq\Pi$.  
We emphasize that this kind of probabilistic noiseless QA is totally different from the classic-like noiseless QA that analogous to the classical amplifier works based on the unambiguous identification of input states and the preparation of desired amplified states \cite{19,20}. Though the successful probability is limited by identification of input states, the gain of classic-like quantum amplifier can be arbitrary high which is not the case for our probabilistic quantum amplifier.

The probabilistic linear amplification of course can be noisy or noiseless when the probabilistic amplification is nonlinear, but we usually do not consider them since they are practically trivial. 

As a summary, three different kinds of important QA, i.e., linear deterministic noisy QA, nonlinear deterministic noiseless QA and linear probabilistic noiseless QA are identified and discussed in detail. For deterministic QA, the linearity of the amplifier is incompatible with the noiseless amplification, while it is not a problem for probabilistic QA. The clear relation between this three different kinds of quantum amplification is shown in Fig. 1.

\section{\bf Phase-preserving amplification of quantum states}
Phase-preserving QA is of theoretical interest and of practical importance \cite{1,2,4,6,7,21,22,23}, especially for the phase-preserving quantum amplification of Gaussian states in phase space.

\subsection{\bf Definition of phase-preserving quantum amplification} 
The phase-preserving QA is the amplification of states with respect to two canonical observables $\hat{X}_{1}$ and $\hat{X}_{2}$, e.g., position $\hat{Q}$ and momentum $\hat{P}$. For any state $\rho_{i}\subseteq\Pi$, the phase-preserving quantum amplifier amplifies the state $\rho_{i}$ into the corresponding amplified  state $\sigma_{i}\subseteq\Xi$ with $\mathrm{Tr}(\hat{X}_{1}\sigma_{i})=g_{i}\mathrm{Tr}(\hat{X}_{1}\rho_{i})$ and $\mathrm{Tr}(\hat{X}_{2}\sigma_{i})=g_{i}\mathrm{Tr}(\hat{X}_{2}\rho_{i})$. If $g_{i}$ is the same for any state $\rho_{i}$, then it is a linear phase-preserving quantum amplifier, otherwise it is non-linear.

\begin{figure}[tbp]
\centering
\includegraphics[scale=0.15]{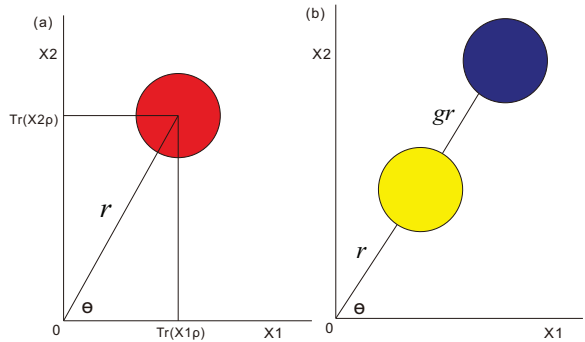}
\caption{(Color Online) {\it Illustration of the distribution of quantum state $\rho$ in $(\hat{X}_{1},\hat{X}_{2})$ space and phase-preserving quantum amplification}. (a) The distribution of quantum state $\rho$ in $(\hat{X}_{1},\hat{X}_{2})$ space. For the simplicity of figure, we assume here that the distribution is Gaussian. The position of distribution totally determined by amplitude $r$ and phase $\theta$. (b) Phase-preserving amplification of quantum state $\rho$. The amplitude is amplified by $g$-fold while the phase stays unchanged.}
\end{figure}

We can construct an operator $\hat{a}=\hat{X}_{1}+i\hat{X}_{2}$ which is not Hermitian. Though $\hat{a}$ is not an observable, the phase-preserving amplification with respect to observables $\hat{X}_{1}$ and $\hat{X}_{2}$ can be regarded as amplification with respect to $\hat{a}$ since $\mathrm{Tr}(\hat{a}\sigma_{i})=g_{i}\mathrm{Tr}(\hat{a}\rho_{i})$. Analogous to phase space based on position $\hat{Q}$ and momentum $\hat{P}$, the general phase space based on canonical observables $\hat{X}_{1}$ and $\hat{X}_{2}$ can be defined. Any quantum state $\rho$ can be represented by its distribution in $(\hat{X}_{1},\hat{X}_{2})$ space that denoted by 
$G_{\rho}(x_{1},x_{2})$. The marginal distribution of $G_{\rho}(x_{1},x_{2})$ can be given in the form
\begin{equation}
\begin{split}
\int dx_{2}G_{\rho}(x_{1},x_{2})&=G_{\rho}(x_{1}),\\
\int dx_{1}G_{\rho}(x_{1},x_{2})&=G_{\rho}(x_{2}),
\end{split}
\end{equation}
where we have assumed that $\hat{X}_{1}$ and $\hat{X}_{2}$ are continuous observables and $G_{\rho}(x_{1})\equiv\langle x_{1}|\rho|x_{1}\rangle$ , $G_{\rho}(x_{2})\equiv\langle x_{2}|\rho|x_{2}\rangle$ are distribution of $\rho$ in the representation of $\hat{X}_{1}$, $\hat{X}_{2}$ respectively. When $\hat{X}_{1}=\hat{Q}$ and $\hat{X}_{2}=\hat{P}$, the distribution $G_{\rho}(x_{1},x_{2})$ is just the Wigner function $W_{\rho}(q,p)$ in phase space.
The phase-preserving amplification of quantum state $\rho$ can be restated as the amplitude of $\rho$ in $(\hat{X}_{1},\hat{X}_{2})$ space being amplified while the phase being unchanged. The phase stays unchanged after the amplification is the reason that we call it phase-preserving amplification. As it can be seen from Fig. 2, the amplitude of $\rho$ in $(\hat{X}_{1},\hat{X}_{2})$ space is defined as $r=\sqrt{\mathrm{Tr}(\hat{X}_{1}\rho)^{2}+\mathrm{Tr}(\hat{X}_{2}\rho)^{2}}$ and the phase is determined by $\theta=\mathrm{arctan}[\dfrac{\mathrm{Tr}(\hat{X}_{2}\rho)}{\mathrm{Tr}(\hat{X}_{1}\rho)}]$.

In practice, the particular focus is given to the spacial case of $\hat{X}_{1}=\hat{Q}$ and $\hat{X}_{2}=\hat{P}$. The distribution of quantum state $\rho$ is described by the Wigner function defined as \cite{24}
\begin{equation}
W_{\rho}(q,p)=\dfrac{1}{2\pi\hbar}\int_{-\infty}^{+\infty}d\xi \mathrm{exp}(-\dfrac{i}{\hbar}p\xi)\langle q+\dfrac{1}{2}\xi|\rho|x-\dfrac{1}{2}\xi\rangle.
\end{equation}
For any two states $\lbrace\rho_{i},\rho_{j}\rbrace$, their overlap is equal to the overlap of their corresponding distributions in phase space \cite{25}
\begin{equation}
\mathrm{Tr}(\rho_{i}\rho_{j})=2\pi\hbar\int_{-\infty}^{+\infty}dq\int_{-\infty}^{+\infty}dp W_{\rho_{i}}(q,p)W_{\rho_{j}}(q,p).
\end{equation}

\subsection{\bf Phase-preserving quantum amplification of Gaussian states}

Gaussian states play an important role in quantum information and quantum communication \cite{26,27,28}. The phase-preserving QA of Gaussian states is not only of theoretical interest \cite{1,10} but also of practical importance \cite{29,30,31,32,33}. 

A Gaussian state is defined as such a state that its characteristic function, or equivalently its Wigner function, is Gaussian.
The vacuum state, coherent states, squeezed states, and thermal states are typical Gaussian states. The Gaussian state is fully characterized by its first moment $\vec{d}$ and second moment $\vec{\gamma}$ \cite{27,28}. For a single-mode Gaussian state $\rho$, the first moment $\vec{d}=(d_{1},d_{2})$ is defined as
\begin{equation}
d_{i}=\mathrm{Tr}(\hat{X}_{i}\rho),
\end{equation}
where $\hat{X}_{i}$ represent quadrature operators $\hat{Q}$ and $\hat{P}$. The second moment $\vec{\gamma}$, which forms the so-called covariance matrix $\vec{\gamma}=(\gamma_{ij})(i,j=1,2)$, is given by
\begin{equation}
\gamma_{ij}=\mathrm{Tr}(\lbrace\hat{X}_{i},\hat{X}_{j}\rbrace\rho)-2d_{i}d_{j},
\end{equation}
where $\lbrace\hat{X}_{i},\hat{X}_{j}\rbrace\equiv\hat{X}_{i}\hat{X}_{j}+\hat{X}_{j}\hat{X}_{i}$ is the anti-commutator of $\hat{X}_{i}$ and $\hat{X}_{j}$.
For simplicity and convenience, only single-mode Gaussian states are considered here. Results about phase-preserving QA of single-mode Gaussian states can be easily extended into multi-mode case.

We consider the phase-preserving QA of a set of $N$ single-mode Gaussian states $\Pi=\lbrace\rho_{1},\rho_{2},...,\rho_{N}\rbrace$ here. Note that all Gaussian states in set $\Pi$ should be in the same single-mode, otherwise different Gaussian states can be identified because of mode identification. The corresponding set of $N$ amplified Gaussian states is denoted by $\Xi=\lbrace\sigma_{1},\sigma_{2},...,\sigma_{N}\rbrace$. For any state $\rho_{i}\subseteq\Pi$, the phase-preserving QA means the amplification of the first moment 
\begin{equation}
\vec{d}_{\sigma_{i}}=g_{i}\vec{d}_{\rho_{i}},
\end{equation}
where $\vec{d}_{\rho}$ represents the first moment of Gaussian state $\rho$. If the gain $g_{i}$ is the same for all states $\rho_{i}\subseteq\Pi$, then the phase-preserving QA of Gaussian states is linear, otherwise it is nonlinear. For noiseless amplification, the second moments should stay unchanged
\begin{equation}
\vec{\gamma}_{\sigma_{i}}=\vec{\gamma}_{\rho_{i}},   \forall\rho_{i}\subseteq\Pi,
\end{equation}
where $\vec{\gamma}_{\rho}$ represents the second moment of state $\rho$.

The distinguishability of any two Gaussian states can be measured by their distance in phase space. The distance between any two Gaussian states $\rho$ and $\sigma$ in phase space can be written as
\begin{equation}
D(\rho,\sigma)=(\vec{d}_{\rho}-\vec{d}_{\sigma})^{2}.
\end{equation}
The distinguishability of two states $\rho$ and $\sigma$ that is inversely proportional to overlap $\mathrm{Tr}(\rho\sigma)$ is also proportional to their distance $D(\rho,\sigma)$ in phase space. The larger the distance between two Gaussian states, the more the distinguishability of these two states and vice versa. 

For a deterministic quantum amplifier, it has been shown above that the linearity of amplification is incompatible with noiseless amplification. The linear deterministic phase-preserving QA of Gaussian states  can not be noiseless. This can be seen by calculating the distance of two noiseless amplified Gaussian states. Suppose that the amplification is noiseless, the distance between any two amplified Gaussian states $\sigma_{i},\sigma_{j}\subseteq\Xi$ is
\begin{equation}
D(\sigma_{i},\sigma_{j})=(\vec{d}_{\sigma_{i}}-\vec{d}_{\sigma_{j}})^{2}=g^{2}D(\rho_{i},\rho_{j}),
\end{equation}
where $D(\rho_{i},\rho_{j})=(\vec{d}_{\rho_{i}}-\vec{d}_{\rho_{j}})^{2}$ and Eq. (27) is used in the last step above. Obviously, for $g>1$, $D(\sigma_{i},\sigma_{j})>D(\rho_{i},\rho_{j})$, which implies that Gaussian states are more distinguishable after deterministic QA, which is directly conflicted with the physical fact that the distinguishability of any two states in the set $\Pi$ does not increase after deterministic QA.

Of course, the linear probabilistic phase-preserving QA of Gaussian states can be noiseless. The gain of amplification $g$ and the successful probability $p$ is limited by Eq. (20). Furthermore, Eq. (21) is satisfied if the successful probability $p$ is the same for all states $\rho_{i}\subseteq\Pi$. 
Consider the linear probabilistic QA of set of two Gaussian states $\lbrace\rho_{1},\rho_{2}\rbrace$ and its corresponding amplified set of coherent states is $\lbrace\sigma_{1},\sigma_{2}\rbrace$ with gain $g$ of QA. If the QA is noiseless then Eq.(21) becomes
\begin{equation}
f(g)=Cp^{2}+V(1-p)^{2},
\end{equation}
where $C\equiv\mathrm{Tr}(\tau_{1}\tau_{2})$ and $V\equiv\dfrac{\mathrm{Tr}(\upsilon_{1}\upsilon_{2})}{\mathrm{Tr}(\sigma_{1}\sigma_{2})}$ are constants in this case and $f(g)$ is proportional to $g$. The minimum gain of QA $g_{min}$ exists since $\dfrac{d^{2}f(g)}{dp^{2}}=2(C+V)>0$. We obtain the minimum point $p_{0}=\dfrac{V}{V+C}$ by setting $\dfrac{df(g)}{dp}=0$. The $g_{min}$ is thus determined by
\begin{equation}
f(g_{min})=\sqrt{Cp_{0}^{2}+V(1-p_{0})^{2}}.
\end{equation}
For the case of QA of coherent states, $f(g)=e^{(g^{2}-1)D(\rho_{1},\rho_{2})}$ and we have
\begin{equation}
g_{min}=\sqrt{\dfrac{1}{2}\dfrac{\mathrm{ln}[Cp_{0}^{2}lnV(1-p_{0})^{2}]}{D(\rho_{1},\rho_{2})}+1}.
\end{equation}
The existence of minimum gain of amplification $g_{min}$ can be easily understood since the amplified state $\sigma_{i}$ should be distinguished from the initial state $\rho_{i}$. To distinguish the two Gaussian states requires the distinguishability or the distance of two Gaussian states is larger or equal to a threshold value $\epsilon$ below which the two Gaussian states can not be distinguished from each other. The threshold value $\epsilon$ determines the minimum gain of amplification $g_{min}$. The distance between state $\rho_{i}$ and the corresponding amplified state $\sigma_{i}$ depend on the first moment of state $\rho_{i}$ and the gain $g$ of amplification since
\begin{equation}
 D(\sigma_{i},\rho_{i})=[(g-1)\vec{d}_{\rho_{i}}]^{2}.
\end{equation}  
The threshold value $\epsilon$ should be proportional to the minimum distance that defined as $D_{min}\equiv\mathrm{min}\lbrace D(\sigma_{i},\rho_{i}),\forall\rho_{i}\subseteq\Pi\rbrace$, that is
\begin{equation}
\epsilon=\kappa D_{min},
\end{equation}
where $\kappa$ is a constant factor for a definite set of Gaussian states $\Pi$. The minimum gain of amplification can be calculated as
\begin{equation}
g_{min}=\sqrt{\dfrac{\epsilon}{\kappa(\vec{d}_{\rho,min})^{2}}}+1,
\end{equation}
where $\vec{d}_{\rho_{i},min}\equiv\mathrm{min}\lbrace \vec{d}_{\rho_{i}},\forall\rho_{i}\subseteq\Pi\rbrace$ is the minimum first moment of states $\rho_{i}$ in set $\Pi$. In general, the situation is more complicated so that we have to consider the set of equations in the form of Eq.(21).

The deterministic phase-preserving QA of Gaussian states can be noiseless if the requirement of linearity of amplification is relaxed. Suppose that the gain of amplification is dependent on different Gaussian states $\rho_{i}\subseteq\Pi$, that is $\mathrm{Tr}(\hat{a}\sigma_{i})=g_{i}\mathrm{Tr}(\hat{a}\rho_{i})$ with $\hat{a}=\hat{Q}+i\hat{P}$. According to Eq. (10), the distinguishability of any two Gaussian states does not increase after deterministic QA. In other words, the distance does not increase after deterministic noiseless QA, i.e.,
\begin{equation}
D(\rho_{i},\rho_{j})\geq D(\sigma_{i},\sigma_{j}).
\end{equation}
Using the definition of distance, the simple calculation gives 
\begin{equation}
\mathrm{cos}\vartheta_{ij}\geq \dfrac{\sqrt{(g_{i}^{2}-1)(g_{j}^{2}-1)}}{g_{i}g_{j}-1},
\end{equation}
where $\vartheta_{ij}$ is the relative phase of states $\rho_{i}$ and $\rho_{j}$ in phase space. For a deterministic noiseless QA, the set of Gaussian states $\Pi$ and gain of amplification $\lbrace g_{i}\rbrace$ must satisfy Eq. (38). On the other hand, if Eq. (38) is satisfied, we can always construct a deterministic noiseless phase-preserving quantum amplifier that amplifies any Gaussian state $\rho_{i}\subseteq\Pi$ to the corresponding amplified Gaussian state $\sigma_{i}\subseteq\Xi$ physically. We thus obtain the following theorem about deterministic noiseless phase-preserving QA of Gaussian states.

{\bf Theorem:} Suppose the set of $N$ Gaussian states $\Pi=\lbrace\rho_{1},\rho_{2},...,\rho_{N}\rbrace$ is to be amplified and its corresponding phase-preserving amplified set is $\Xi=\lbrace\sigma_{1},\sigma_{2},...,\sigma_{N}\rbrace$ with $\vec{d}_{\rho_{i}}=g_{i}\vec{d}_{\sigma_{i}}$ and $\vec{\gamma}_{\sigma_{i}}=\vec{\gamma}_{\rho_{i}}$ for any Gaussian state $\rho_{i}\subseteq\Pi$. The deterministic noiseless phase-preserving quantum amplifier that amplifies the state $\rho_{i}$ randomly chosen from set $\Pi$ into the corresponding amplified state $\sigma_{i}$ in the set $\Xi$ exists if and only if the states in set $\Pi$ and the gain of amplification satisfy the condition $\mathrm{cos}\vartheta_{ij}\geq \dfrac{\sqrt{(g_{i}^{2}-1)(g_{j}^{2}-1)}}{g_{i}g_{j}-1}$ for any two states $\rho_{i},\rho_{j}\subseteq\Pi$, where $\vartheta_{ij}$ is the relative phase between states $\rho_{i}$ and $\rho_{j}$ in phase space.

There exists a special case that all the final amplified states have the same amplitude in phase space. In this case, $\vec{d}_{\sigma_{i}}^{2}=\vec{d}_{\sigma_{j}}^{2}$ for any two states $\sigma_{i},\sigma_{j}\subseteq\Xi$, which implies that $g_{i}^{2}\vec{d}_{\rho_{i}}^{2}=g_{j}^{2}\vec{d}_{\rho_{j}}^{2}$. Combining this requirement with Eq. (38) finally gives
\begin{equation}
\mathrm{cos}\vartheta_{ij}\geq \sqrt{\dfrac{g_{i}^{2}\vec{d}_{\rho_{i}}^{2}-\vec{d}_{\rho_{j}}^{2}}{(g_{i}^{2}-1)\vec{d}_{\rho_{i}}^{2}\vec{d}_{\rho_{j}}^{2}}}.
\end{equation}

{\bf Corollary:} The deterministic noiseless phase-preserving quantum amplifier that amplifies the Gaussian state $\rho_{i}$ randomly chosen from set $\Pi$ to corresponding amplified Gaussian state $\sigma_{i}$ in the set $\Xi$ with all the same amplitude in phase space exists if and only if the Gaussian states in set $\Pi$ and the gain of amplification satisfy the condition $\mathrm{cos}\vartheta_{ij}\geq \sqrt{\dfrac{g_{i}^{2}\vec{d}_{\rho_{i}}^{2}-\vec{d}_{\rho_{j}}^{2}}{(g_{i}^{2}-1)\vec{d}_{\rho_{i}}^{2}\vec{d}_{\rho_{j}}^{2}}}$ for any two states $\rho_{i},\rho_{j}\subseteq\Pi$, where $\vartheta_{ij}$ is the relative phase between states $\rho_{i}$ and $\rho_{j}$ in phase space.

\subsection{\bf The phase measurement with phase-preserving amplification}
After phase-preserving QA, the phase of Gaussian states stays unchanged. This property can be exploited to improve the precision of phase measurement, for instance, the phase measurement of coherent states using a balanced homodyne detector.

The difference of two photo-detector measurements is obtained in homodyne detection and the output signal is determined by 
\begin{equation}
\hat{n}_{d}=-i(\hat{b}^{\dagger}\hat{c}-\hat{c}^{\dagger}\hat{b}),
\end{equation}
where $\hat{n}_{d}$ is the number difference operator and $\hat{b},\hat{c}$ are annihilation operators \cite{34}. A large amplitude coherent state $|c\rangle$ with fixed phase is used in one input port as reference. When a coherent state $|b\rangle$ is sent into another input port, the mean measured signal at the output is
\begin{equation}
\langle\hat{n}_{d}\rangle=2|b|\cdot|c|\mathrm{sin}\delta, 
\end{equation}
where $\delta$ is the relative phase between coherent states $|b\rangle$ and $|c\rangle$. The variance of signal is calculated as
\begin{equation}
\Delta\hat{n}_{d}=\sqrt{|b|^{2}+|c|^{2}}.
\end{equation}
The sensitivity of the measured phase, according to error transfer formula, becomes
\begin{equation}
\Delta\delta=\dfrac{\Delta\hat{n}_{d}}{|\partial\langle\hat{n}_{d}\rangle/\partial\delta|}=\dfrac{\sqrt{1+(|c|/|b|)^{2}}}{2|c|\mathrm{cos}\delta}.
\end{equation}

Suppose that the input coherent state $|b\rangle$ is randomly chosen from a definite set of coherent states. A probabilistic noiseless quantum amplifier can be applied to amplify input coherent state $|b\rangle$ before detection. According to Eq. (41) and Eq. (43), not only the mean output signal is enhanced but also the precision of measured phase is improved in this case. Moreover, if this set of coherent states satisfy the condition of deterministic noiseless QA, a deterministic noiseless quantum amplifier can be designed to amplify the input coherent state noiselessly and determinately. 

\subsection{\bf Deterministic amplification in noisy circumstance}
The distinguishabilty of two quantum states decreases in general after deterministic QA. However, the situation may be different in noisy circumstance. The distinguishability of two amplified states through the noisy channel may be larger than the two states through the same noisy channel without amplification.

In order to see it explicitly, suppose that the two states $\rho_{1},\rho_{2}$ are to be amplified and $\sigma_{1},\sigma_{2}$ are the two corresponding amplified states. The noisy channel can be described by superoperator $\hat{V}(t)$ so that the state evolution of quantum system in noisy channel is $\rho(t)=\hat{V}(t)[\rho(0)]$ \cite{35}. In general, the distance of two states decreases monotonously in noisy channel because of the information dissipation of quantum system $D(\rho_{1}(t),\rho_{2}(t))\leq D(\rho_{1}(0),\rho_{2}(0))$. The decay rate of distance can de defined as \cite{36}
\begin{equation}
\chi(\rho_{1}(t),\rho_{2}(t))=\dfrac{d}{dt}D(\rho_{1}(t),\rho_{2}(t)).
\end{equation}
$\chi(\rho_{1}(t),\rho_{2}(t))\leq 0$ implies the distinguishability of two states decreases with time in noisy channel while in non-Markovian channel $\chi(\rho_{1}(t),\rho_{2}(t))>0$ \cite{37,38}. The decay rate of two amplified quantum states in the same noisy channel can be defined similarly. The decay rate of distance depends on the initial states which means the possible different decay rate for states and amplified states. It is thus possible that the distinguishability of two amplified states may be larger than the states without amplification through the same noisy channel. 

Besides, the detector is not truly ideal in practice. The unavoidable noises in detector will lower our precision of distinguish quantum states. For an imperfect detector, the amplified states may be more distinguishable than states without amplification.

\section{\bf Summary}
In this paper, using the language of quantum state transformation, we give a general framework of quantum state amplification systematically for the first time. Based on explicit definition of amplification of quantum states, we formulate the amplification of set of quantum states as transformation of quantum states. Three different kinds of important QA, i.e., deterministic noisy QA, probabilistic noiseless QA, and deterministic noiseless QA, are identified and discussed in detail. For deterministic QA, we show that the linearity of amplification is incompatible with the noiseless amplification while it is not a problem for probabilistic QA. For probabilistic noiseless QA, we discussed the connection between the gain of amplification and the successful probability. We obtain a generally valid relation between the gain of amplification and the successful probability for the case in which successful probability is the same for all quantum states to be amplified.
Particular interest is focused on phase-preserving QA of Gaussian states that has been shown of theoretical interest and of practical importance in quantum information and quantum communication. Using the concept of distance between Gaussian states, the theorem of deterministic noiseless QA of Gaussian states with explicit formula is obtained for the first time. Also, the application of noiseless QA in phase measurement is given and deterministic QA in noisy circumstance is discussed. Though the distinguishability of two quantum states decreases after deterministic amplification in general, the situation may be very different in noisy circumstance. Our discussion of quantum state amplification viewed as quantum state transformation provides a general framework to discuss quantum amplification. The results of quantum state amplification we obtained not only enrich the research of quantum amplification but also may be helpful for further applications of quantum amplification.

{\bf Acknowledgements} 

The authors thank Sheng Liu for helpful discussions. This work is supported by National Natural Science Foundation of China (No. 61275122, No. 61590932) and Strategic Priority Research Program (B) of CAS (No. XDB01030200).

\end{document}